\documentclass[preprint,3p,times]{elsarticle}

\usepackage{lineno,hyperref}
\usepackage[T1]{fontenc}
\usepackage{comment}
\usepackage{amsmath}
\usepackage{amssymb}
\usepackage{graphicx}
\usepackage{epsfig}
\usepackage{latexsym}
\usepackage{pstricks}
\usepackage{pst-plot}
\usepackage{amsmath, amsthm, amsfonts, amssymb}
\usepackage{appendix}
\usepackage[bitstream-charter]{mathdesign}
\modulolinenumbers[5]

\journal{Journal of \LaTeX\ Templates}

\newtheorem{theorem}{Theorem}[section]
\newtheorem{lemma}[theorem]{Lemma}

\begin{document}

\begin{frontmatter}

\title{Gaussian ensembles distributions from mixing quantum systems}
\author[iflp]{Ignacio S. Gomez\corref{cor1}}
\ead{nachosky@fisica.unlp.edu.ar}
\author[iflp]{M. Portesi}
\ead{portesi@fisica.unlp.edu.ar}

\cortext[cor1]{Corresponding author}
\address[iflp]{IFLP, UNLP, CONICET, Facultad de Ciencias Exactas, Calle 115 y 49, 1900 La Plata, Argentina}

\begin{abstract} In the context of the mixing dynamical systems we present a derivation of the Gaussian ensembles distributions from mixing quantum systems having a classical analog that is mixing. We find that mixing factorization property is satisfied for the mixing quantum systems expressed as a factorization of quantum mean values. For the case of the kicked rotator and in its fully chaotic regime, the factorization property links decoherence by dephasing with Gaussian ensembles in terms of the weak limit, interpreted as a decohered state. Moreover, a discussion about the connection between random matrix theory and quantum chaotic systems, based on some attempts made in previous works and from the viewpoint of the mixing quantum systems, is presented.

\end{abstract}

\begin{keyword}
Gaussian ensembles \sep Mixing \sep Quantum Mixing \sep Weak limit
\end{keyword}

\end{frontmatter}

\nolinenumbers

\section{Introduction}
Gaussian ensemble theory emerged from the study of complex nuclei and long lived resonance states in the 1950s by Wigner \cite{Wigner gaussian}, and later by Dyson \cite{Dyson}.
Wigner's central idea was that for quantum systems with many degrees of freedom like a heavy nucleus, one can assume that the Hamiltonian matrix elements in a typical basis can be treated as independent Gaussian random numbers. The main prediction of this approach is that the statistical distribution of spacings between adjacent energy levels obeys universal distributions which define the \emph{Gaussian Orthogonal Ensemble}, the \emph{Gaussian Unitary Ensemble} and the \emph{Gaussian Symplectic Ensemble}, if the Hamiltonian is invariant under an orthogonal, unitary or symplectic transformation, respectively. In 1984 Bohigas, Gianonni and Schimt \cite{Boh84} formulated their celebrated statement (briefly named as BGS conjecture) concerning quantum chaotic systems:
\textit{Spectral of time-reversal invariant systems whose classical analogue are $K$-systems show the same statistical properties as predicted by Gaussian Orthogonal Ensembles}. Moreover, Gaussian ensembles proved to be powerful tool to study statistical properties in many applications
\cite{apps1,apps2, apps3,apps4}.



Ergodic hierarchy (EH) classifies the chaos of dynamical systems according to the decay of correlations between subsets of the phase space for large times. $K$--systems correspond to the Kolmogorov level of the EH.
Related to this, in \cite{0, NACHOSKY MARIO} a quantum extension of the EH was proposed, called the quantum ergodic hierarchy (QEH), which expresses the decay of correlations between states and observables in the asymptotic limit. In \cite{NACHOSKY MARIO, casati model} the chaotic behaviors of the Casati-Prosen model \cite{casati verdadero} and the kicked rotator \cite{stockmann, haake} were interpreted in terms of the quantum mixing level.

Using the idea of ranking chaos looking at the decay of correlations as in \cite{0, NACHOSKY MARIO}, we perform two previous steps to study the Gaussian ensembles from the quantum mixing level. First, we deduce the mixing factorization property which expresses the classical mean value of a product of observables as a product of mean values. Second, we obtain the quantum analogue of this property in the classical limit and apply it to deduce the Gaussian ensembles. In this way, the contribution of the present paper is to show that Gaussian ensembles are a natural consequence of quantum mixing correlations in the classical limit.




\section{Gaussian Ensembles}
Gaussian ensembles describe how the Hamiltonian matrix elements are distributed in a chaotic quantum system when the details of interactions can be neglected. The surprising prediction capability of the GE lies in the simplicity of the assumptions. If we have a quantum system having a $N\times N$--dimensional Hamiltonian,
in addition to normalization, the two conditions for the probability density function $P(H_{11},H_{12},\ldots,H_{NN})$ of the Hamiltonian matrix elements $H_{ij}$ which define the Gaussian ensembles are (see, for instance, \cite[pp. 73, 74]{stockmann} and \cite[p. 62]{haake})
\begin{eqnarray}\label{1}
P(H_{11},H_{12},\ldots,H_{NN})=P(H_{11})P(H_{12})\cdots P(H_{NN})
\end{eqnarray}
and
\begin{eqnarray}\label{2}
P(H_{11}^{\prime},H_{12}^{\prime},\ldots,H_{NN}^{\prime})=P(H_{11},H_{12},\ldots,H_{NN})
\end{eqnarray}
where the transformed Hamiltonian $\hat{H}^{\prime}$ is obtained from the original one $\hat{H}$ by an orthogonal, unitary or symplectic transformation according to corresponding Gaussian ensemble. Eq. \eqref{2} simply represents the invariance of the density probability $P(H_{11},H_{12},\ldots,H_{NN})$ under an orthogonal, unitary or simplectic transformation.
Eq. \eqref{1} expresses that in the fully chaotic regime of a classically chaotic quantum system, the details of the interactions are not relevant so the Hamiltonian can be replaced by a matrix whose elements are uncorrelated.

\section{Mixing correlations}

In ergodic theory, the decay of correlations is the most important feature for the validity of the statistical description
because different regions of phase space become statistically independent when they are enough separated in time. More precisely, if one has a dynamical system $(\Gamma,\mu,\Sigma,\{T_t\})$ where $\Gamma$ is the phase space, $\mu:\Sigma \rightarrow [0,1]$ is a normalized measure on $\Sigma$, and $\{T_t\}_{t\in J}$ is a semigroup of preserving--measure transformations ($J$ is typically the real numbers), then the EH correlation between two subsets $A,B\subseteq\Gamma$ separated a time $t$ is mathematically expressed as
\begin{equation}\label{correlation}
C(T_tA,B)=\mu(T_tA \cap B)-\mu(A)\mu(B)
\end{equation}
The mixing level of the EH corresponds to the situation when
\begin{equation}\label{mixing1}
\lim_{t\rightarrow\infty}C(T_tA,B)=0
\end{equation}
for all $A,B\subseteq\Gamma$. Several examples like Sinai billiards, Brownian motion, chaotic maps, belong to the mixing level satisfying the eq. \eqref{mixing1}.
The Frobenius-Perron operator $P_t$ associated to the transformation
$T_t$ gives the evolution of any distribution $f$ (i.e. $f:\Gamma\rightarrow[0,\infty]$ with $||f||=1$) by means of
\begin{equation}\label{perron}
\int_{T_{-t}A}f(q,p)dqdp=\int_Af(q,p)dqdp  \ \ \ \  \ \ \ \ \forall \ A\subseteq\Gamma \ \ , \ \ \forall \ t\in J
\end{equation}
where $(q,p)\in\Gamma$. When $P_t$
has a fixed point $f_{\ast}$, i.e. $P_tf_{\ast}=f_{\ast}$, the following relevant property of mixing systems can be deduced.
\begin{lemma}\emph{(Factorization property)}\label{lemma1}
Let $f_{\ast}$ be a normalized distribution which is a fixed point of the Frobenius--Perron operator $P_t$. If $1_{A_1}, 1_{A_2},\ldots,1_{A_n}:\Gamma\rightarrow \mathbb{R}$ are the $n$ characteristic functions of $n$ subsets $A_1,\ldots,A_n\subseteq\Gamma$ then
\begin{equation}\label{lemma1-4}
\begin{split}
\int_\Gamma f_{*}(q,p) 1_{A_1}(q,p)\cdots1_{A_n}(q,p)dqdp=\left(\int_\Gamma f_{*}(q,p) 1_{A_1}(q,p)dqdp \right)\cdots\left(\int_\Gamma f_{*} 1_{A_n}(q,p)dqdp \right)
\end{split}
\end{equation}
\end{lemma}
\noindent Lemma \ref{lemma1} implies that the average of a product weighted by a distribution $f_*(q,p)$ (that is a fixed point of $P_t$) can be factorized in the corresponding product of the averages weighted by the same $f_*(q,p)$. The ``factorization property" of Eq. (\ref{lemma1-4}) is essential in order to obtain the  Gaussian ensembles, we explore its consequences in the context of quantum mixing correlations.

\section{Quantum mixing correlations}

A quantum counterpart of mixing correlation of Eq. (\ref{mixing1}) was derived in \cite{0}. More precisely, in the quantum version of Eq. (\ref{mixing1}) we have a decay correlation between states and observables rather than between subsets of phase space given by
\begin{equation}\label{qcorrelation}
\begin{split}
C(\hat{\rho}(t),\hat{O})=\langle\hat{O}\rangle_{\hat{\rho}(t)}-\langle\hat{O}\rangle_{\hat{\rho}_{*}}
\end{split}
\end{equation}
where the role played in \eqref{correlation} by the subsets $A,B$ is now played by the state $\hat{\rho}(t)$ and the observable $\hat{O}$, with $\hat{\rho}(t)$ being any quantum state $\hat{\rho}$ at time $t$. The state $\hat{\rho}_{\ast}$ is the \emph{weak limit} of $\hat{\rho}$ given by the quantum mixing level of the quantum version of the ergodic hierarchy, i.e. the quantum ergodic hierarchy (QEH)
\begin{equation}\label{qmixing1}
\begin{split}
\lim_{t\rightarrow\infty}C(\hat{\rho}(t),\hat{O})=0  \ \ \ \ \ \ \ \ \ \ \ \ \textrm{for} \ \textrm{all} \ \textrm{observable} \ \  \hat{O}
\end{split}
\end{equation}
We can see the similarity between the mixing correlation and its quantum version (Eqns. \eqref{mixing1} and \eqref{qmixing1}), that is, one can obtain one correlation from the other simply exchanging
$C(T_tA,B)$ by $C(\hat{\rho}(t),\hat{O})$ and vice versa. Eq. \ref{qmixing1} describes the relaxation of any quantum state $\hat{\rho}$ with a weak limit $\hat{\rho}_{\ast}$ where the relaxation is understood in the sense of the quantum mean values, i.e. the decoherence of observables \cite{vhove, Omnes}. We show that
the weak limit $\hat{\rho}_{*}$ is the quantum analogue of the distribution $f_{\ast}$ of Lemma \ref{lemma1}. This is the content of the following result.
\begin{lemma}\label{lemma2}
The state $\hat{\rho}_{\ast}$ is a fixed point of the evolution operator $\hat{U}_t=e^{-it\frac{\hat{H}}{\hbar}}$ where $\hat{H}$ is the Hamiltonian of the quantum system, i.e. $\hat{U}_t\hat{\rho}_{\ast}\hat{U}_t^{\dag}=\hat{\rho}_{\ast}$.
\end{lemma}
\noindent In order to establish a quantum version of Lemma \ref{lemma1} we recall some properties of the Weyl symbol and the Wigner function.
If $\hat{A}$ is an operator then its Weyl symbol $\widetilde{W}_{\hat{A}}$ is a distribution function over phase space defined by \cite{Wigner, Symb}
\begin{eqnarray}\label{wigner1}
\widetilde{W}_{\hat{A}}(q,p)=\int_{\mathbb{R}}\left\langle q+\frac{\Delta}{2}\left|\,\hat{A}
\,\right|q-\frac{\Delta}{2}\right\rangle e^{-i\frac{p\Delta}{\hbar}}d\Delta
\end{eqnarray}
In particular, if $1_E(q,p)$ is the characteristic function of a subset $E$ of $\Gamma$ we will use the Weyl symbol of $\hat{\pi}_E$, with
\begin{eqnarray}\label{weylcharacteristic}
\widetilde{W}_{\hat{\pi}_E}(q,p)=1_E(q,p)   \ \ \ \ \ \forall \ (q,p)\in\mathbb{R}^2
\end{eqnarray}
The Wigner function $W_{\hat{A}}$ is defined by means of the Weyl symbol as
\begin{eqnarray}\label{wigner1bis}
W_{\hat{A}}(q,p)=\frac{1}{h}\widetilde{W}_{\hat{A}}(q,p)
\end{eqnarray}
A relevant property of the Wigner function is that it allows to express any quantum mean value as an integral in phase space \cite{Wigner}, in the form
\begin{eqnarray}\label{wigner8}
\langle\hat{O}\rangle_{\hat{\rho}}=
\int_{\mathbb{R}^2} dqdp \ W_{\hat{\rho}}(q,p)\widetilde{W}_{\hat{O}}(q,p)
\end{eqnarray}
For the Weyl symbol of a product of operators, it can be shown that the following expansion is fulfilled
\begin{eqnarray}\label{wigner4}
\widetilde{W}_{\hat{A}\hat{B}}(q,p)=\widetilde{W}_{\hat{A}}(q,p)\widetilde{W}_{\hat{B}}(q,p)+\textrm{O}(\hbar)
\end{eqnarray}
An important property that we will use can be deduced by the definition of Weyl symbol in the classical limit $\hbar\rightarrow0$.
\begin{lemma}\label{lemaweyl}
Let $\widetilde{W}_{\hat{A}}(q,p)$ be the Weyl symbol of an operator $\hat{A}$. Then in the classical limit of $\hbar\rightarrow0$ the Weyl symbol of $\hat{A}(-t)=\hat{U}_t^{\dag}\hat{A}\hat{U}_t$ is
$\widetilde{W}_{\hat{A}}(q(t),p(t))$, where $(q(t),p(t))=(T_tq,T_tp)$ and $T_t$ is the classical evolution given by Hamilton equations. That is,
\begin{eqnarray}\label{lemaweyl1}
\widetilde{W}_{\hat{U}_t^{\dag}\hat{A}\hat{U}_t}(q,p)=\widetilde{W}_{\hat{A}}(q(t),p(t))  \ \ \ \ \ \ \ \ \ \ \ \ \forall \ (q,p)\in \mathbb{R}^2 \ , \ \forall \ t\in\mathbb{R}
\end{eqnarray}
\end{lemma}
For quantum mixing correlations the following property in phase space is a consequence of Lemmas \ref{lemma2} and \ref{lemaweyl}.
\begin{lemma}\label{corollary}
The Wigner distribution $W_{\hat{\rho}_{\ast}}(q,p)$ is a fixed point of
the Frobenius-Perron operator $P_t$ associated with the classical evolution $T_t$ given by Hamiltonian equations.
\end{lemma}
Now joining the previous Lemmas \ref{lemma1}, \ref{lemma2}, \ref{lemaweyl} and \ref{corollary} we show a quantum analogue of the factorization property (i.e. Eq. \eqref{lemma1-4}). This is one of the main results of the present contribution.
\begin{theorem}\emph{(Quantum factorization property)}\label{theorem}
Assume one has a mixing quantum system, i.e., the correlation $C(\hat{\rho}(t),\hat{O})$ of $\hat{\rho}(t)$ with any observable $\hat{O}$
vanishes for $t\rightarrow\infty$. Then, for a set of observables $\hat{O}_1,\ldots,\hat{O}_n$ when $\hbar\rightarrow0$ one has
\begin{equation}\label{teo2-1}
\langle\hat{O}_1\cdots\hat{O}_n\rangle_{\hat{\rho}_{*}}=\langle\hat{O}_1\rangle_{\hat{\rho}_{*}}\cdots\langle\hat{O}_n\rangle_{\hat{\rho}_{*}}
\end{equation}
\begin{proof}
In principle, the Wigner property of Eq. \eqref{wigner8} applied to the product $\hat{O}_1\cdots\hat{O}_n$ and $\hat{\rho}_{\ast}$ gives us
\begin{equation}\label{lemma1-12}
\begin{split}
\langle\hat{O}_1\cdots\hat{O}_n\rangle_{\hat{\rho}_{*}}=\int_{\mathbb{R}^2}W_{\hat{\rho}_{\ast}}(q,p)\widetilde{W}_{\hat{O}_1\cdots\hat{O}_n}(q,p)\ dqdp
\end{split}
\end{equation}
Applying several times Eq. \eqref{wigner4} on $\hat{O}_1\cdots\hat{O}_n$ we have
\begin{equation}\label{lemma1-13}
\widetilde{W}_{\hat{O}_1\cdots\hat{O}_n}(q,p)=\widetilde{W}_{\hat{O}_1}(q,p)\cdots\widetilde{W}_{\hat{O}_n}(q,p)+\textrm{O}(\hbar)
\end{equation}
From Eqns. \eqref{lemma1-12}, \eqref{lemma1-13} and since $\int_{\mathbb{R}^2}W_{\hat{\rho}_{\ast}}(q,p)dqdp=1$ it follows that
\begin{equation}\label{lemma1-14}
\langle\hat{O}_1\cdots\hat{O}_n\rangle_{\hat{\rho}_{*}}=
\int_{\mathbb{R}^2}W_{\hat{\rho}_{\ast}}(q,p)\widetilde{W}_{\hat{O}_1}(q,p)\cdots\widetilde{W}_{\hat{O}_n}(q,p)\ dqdp+\textrm{O}(\hbar)
\end{equation}
Then, in the classical limit $\hbar\rightarrow0$ we can neglect terms of order $\textrm{O}(\hbar)$ so \eqref{lemma1-14} becomes
\begin{equation}\label{lemma1-14bis}
\langle\hat{O}_1\cdots\hat{O}_n\rangle_{\hat{\rho}_{*}}=
\int_{\mathbb{R}^2}W_{\hat{\rho}_{\ast}}(q,p)\widetilde{W}_{\hat{O}_1}(q,p)\cdots\widetilde{W}_{\hat{O}_n}(q,p)\ dqdp   \ \ \ \ \ \ \ \ when \ \ \ \ \ \ \ \ \hbar\rightarrow0
\end{equation}
We can expand $\widetilde{W}_{\hat{O}_1},\ldots,\widetilde{W}_{\hat{O}_n}$ as linear combinations of characteristic functions.

That is, $\widetilde{W}_{\hat{O}_1}(q,p)=\sum_j\alpha_{1j}1_{C_{1j}}(q,p)$,\ldots,$\widetilde{W}_{\hat{O}_n}(q,p)=\sum_l\alpha_{nl}1_{C_{nl}}(q,p)$. Then we have
\begin{equation}\label{lemma1-15}
\begin{split}
&\langle\hat{O}_1\cdots\hat{O}_n\rangle_{\hat{\rho}_{*}}=
\int_{\mathbb{R}^2}W_{\hat{\rho}_{\ast}}(q,p)\left(\sum_j\alpha_{1j}1_{C_{1j}}(q,p)\cdots\sum_l\alpha_{nl}1_{C_{nl}}(q,p)\right) dqdp\nonumber\\
&=\sum_j\alpha_{1j}\cdots\sum_l\alpha_{nl}\int_{\mathbb{R}^2}W_{\hat{\rho}_{\ast}}(q,p)1_{C_{j1}}(q,p)\cdots1_{C_{nl}}(q,p)dqdp
\end{split}
\end{equation}
Now since $W_{\hat{\rho}_{\ast}}(q,p)$ is a fixed point of $P_t$, as shown in Lemma \ref{lemma2}, then we can apply the factorization property of Lemma \ref{lemma1} to the integral in the right hand of Eq. \eqref{lemma1-15}:
\begin{equation}\label{lemma1-19}
\begin{split}
&\int_{\mathbb{R}^2}W_{\hat{\rho}_{\ast}}1_{C_{1j}}(q,p)\cdots1_{C_{nl}}(q,p)dqdp=
\int_{\mathbb{R}^2}W_{\hat{\rho}_{\ast}}1_{C_{1j}}(q,p)dqdp\cdots
\int_{\mathbb{R}^2}W_{\hat{\rho}_{\ast}}1_{C_{nl}}(q,p)dqdp
\end{split}
\end{equation}
This implies that
\begin{eqnarray}\label{lemma1-19}
&\sum_j\alpha_{1j}\cdots\sum_l\alpha_{nl}\int_{\mathbb{R}^2}W_{\hat{\rho}_{\ast}}(q,p)1_{C_{1j}}(q,p)\cdots1_{C_{nl}}(q,p)dqdp\nonumber\\
&=\int_{\mathbb{R}^2}W_{\hat{\rho}_{\ast}}(q,p)\sum_j\alpha_{1j}1_{C_{1j}}(q,p)dqdp\cdots
\int_{\mathbb{R}^2}W_{\hat{\rho}_{\ast}}(q,p)\sum_l\alpha_{nl}1_{C_{nl}}(q,p)dqdp\nonumber\\
&=\int_{\mathbb{R}^2}W_{\hat{\rho}_{\ast}}(q,p)\widetilde{W}_{\hat{O_1}}(q,p)dqdp\cdots
\int_{\mathbb{R}^2}W_{\hat{\rho}_{\ast}}(q,p)\widetilde{W}_{\hat{O_n}}(q,p)dqdp=\nonumber\\
&\langle\hat{O}_1\rangle_{\hat{\rho}_{\ast}}\cdots\langle\hat{O}_n\rangle_{\hat{\rho}_{\ast}}
\end{eqnarray}
which ends the proof.
\end{proof}
\end{theorem}
Theorem \ref{theorem} expresses the quantum version of mixing correlations in the classical limit $\hbar\rightarrow0$.
\section{Gaussian ensembles by means of mixing quantum systems}
The manifestation of chaotic aspects in quantum systems is possible within characteristic timescales $t\lesssim\tau$ (with $\tau \propto\hbar^{-\alpha}$ in the regular case being $\alpha$ proportional to the phase space dimension, and $\tau \propto-\log{\hbar}$ in the chaotic case). In these timescales, the semiclassical and quantum descriptions overlap with the particularity that in the logarithmic timescale $-\log{\hbar}$ the statistical predictions of the Gaussian ensembles are displayed \cite{stockmann, haake, casati libro, chirikov izrailev}. Moreover, within the logarithmic timescale it is expected that the states contain statistical properties as the randomness and invariance conditions [Eqs.\eqref{1} and \eqref{2}], and expressed in terms of quantum correlations. This motivates the following connection between Gaussian ensembles and mixing quantum systems.

As we have shown in Sec. 4, the quantum correlations of mixing quantum systems are contained in the weak limit $\hat{\rho}_*$ which is representative of the quantum system in the asymptotic limit $t\rightarrow\infty$. And since the logarithmic timescale imposes that $t\leq-\log\hbar$ then the asymptotic limit can be guaranteed in the classical limit for $\hbar$ vanishingly small.

The following lemma constitutes a useful tool in order to deduce the Gaussian ensembles within the mixing quantum formalism.

\begin{lemma}\label{lema GE}
Assume one has a quantum system $S$ subject to a Hamiltonian $\hat{H}$  with Hamiltonian matrix elements having a density probability function $P(H_{11},H_{12},\ldots,H_{NN})$.
Let $P_{11}(H_{11}),P_{12}(H_{12}),\ldots,P_{NN}(H_{NN})$ be the marginals of \mbox{$P(H_{11},H_{12},\ldots,H_{NN})$} with respect to the variables $H_{11},H_{12},\ldots,H_{NN}$.
Then
for each set of values $P(H_{11},H_{12},\ldots,H_{NN})$, $P_{11}(H_{11}),P_{12}(H_{12}),\ldots,P_{NN}(H_{NN})$ and for $dH_{11},dH_{12},\ldots,dH_{NN}$
sufficiently small, there exist projectors \mbox{$\hat{\pi}(H_{11},H_{12},\ldots,H_{NN})$}, $\hat{\pi}_{11}(H_{11},H_{12},\ldots,H_{NN}),\hat{\pi}_{12}(H_{11},H_{12},\ldots,H_{NN})$,$\ldots$, \mbox{$\hat{\pi}_{NN}(H_{11},H_{12},\ldots,H_{NN})$}, and a weak limit $\hat{\rho}_{*}(H_{11},H_{12},\ldots,H_{NN})$ such that
\begin{eqnarray}\label{GE-1}
\langle \hat{\pi}_{ij}(H_{11},H_{12},\ldots,H_{NN})\rangle_{\hat{\rho}_{*}(H_{11},H_{12},\ldots,H_{NN})}=P_{ij}(H_{ij})dH_{ij} \ \ \ \ \ \ \ \ \ \ \forall \ i,j=1,\ldots,N
\end{eqnarray}
and
\begin{eqnarray}\label{GE-2}
\langle \hat{\pi}(H_{11},H_{12},\ldots,H_{NN})\rangle_{\hat{\rho}_{*}(H_{11},H_{12},\ldots,H_{NN})}=P(H_{11},H_{12},\ldots,H_{NN})dH_{11}dH_{12}\cdots dH_{NN}
\end{eqnarray}
Moreover, in the classical limit $\hbar\rightarrow0$, the projector $\hat{\pi}(H_{11},H_{12},\ldots,H_{NN})$ can be expressed in terms of \\
$\hat{\pi}_{11}(H_{11},H_{12},\ldots,H_{NN})$,
$\hat{\pi}_{12}(H_{11},H_{12},\ldots,H_{NN}),\ldots,\hat{\pi}_{NN}(H_{11},H_{12},\ldots,H_{NN})$ as
\begin{eqnarray}\label{GE-3}
\hat{\pi}=\hat{\pi}_{11}\hat{\pi}_{12}\cdots\hat{\pi}_{NN} .
\end{eqnarray}
\end{lemma}
\noindent Two remarks can be made regarding this lemma. First,
the product $P_{ij}(H_{ij})dH_{ij}$
gives the probability that the $ij$--th Hamiltonian matrix element belongs to the interval
$(H_{ij}, H_{ij}+dH_{ij})$, and a similar statement follows for the product $P(H_{11},H_{12},\ldots,H_{NN})dH_{11}dH_{12}\cdots dH_{NN}$ which is the joint probability of the former. Besides, due to previous remark and since $\hat{\pi}$, $\hat{\pi}_{11}\hat{\pi}_{12},\ldots,\hat{\pi}_{NN}$ are projectors, then Eqs.\eqref{GE-1} and \eqref{GE-2} express a sort of Born rule \cite{var} performed by means of weak limit states.
Lemma 5.1 allows one to obtain the randomness and invariance conditions that define Gaussian ensembles. This is the content of the following theorem.

\begin{theorem}\emph{(Gaussian ensembles distributions from mixing quantum systems)}\label{teo GE}
\begin{itemize}
  \item[$(i)$] Assuming that $S$ is a mixing quantum system, then in the classical limit $\hbar\rightarrow0$ one obtains the \textsc{randomness condition}
\begin{eqnarray}\label{GE-4}
P(H_{11},H_{12},\ldots,H_{NN})=P_{11}(H_{11})P_{12}(H_{12})\cdots P_{NN}(H_{NN})\nonumber
\end{eqnarray}
\item[$(ii)$] Let us consider the transformed variables $H_{11}^{\prime},H_{12}^{\prime},\ldots,H_{NN}^{\prime}$ corresponding to
 $H_{11},H_{12},\ldots,H_{NN}$ through the change of variables $\hat{H}^{\prime}=\hat{U}\hat{H}\hat{U}^{\dag}$ where $\hat{U}$ stands for the transpose, complex transpose, or dual of $\hat{U}$ if $\hat{U}$ is orthogonal, unitary, or simplectic,
respectively. Let $P(H_{11}^{\prime},H_{12}^{\prime},\ldots,H_{NN}^{\prime})$ be the transformed probability density function of $P(H_{11},H_{12},\ldots,H_{NN})$. One obtains the \textsc{invariance condition}
\begin{eqnarray}\label{GE-5}
P(H_{11}^{\prime},H_{12}^{\prime},\ldots,H_{NN}^{\prime})=P(H_{11},H_{12},\ldots,H_{NN})\nonumber
\end{eqnarray}
\end{itemize}
\begin{proof}

\begin{itemize}
  \item[$(i)$] If one applies the quantum factorization property (Theorem \ref{theorem}) and the Lemma \ref{lemma1} to the projectors $\hat{\pi}_{ij}(H_{11},H_{12},\ldots,H_{NN})$, in the classical limit $\hbar\rightarrow0$ one obtains
\begin{eqnarray}\label{GE-6}
\langle \hat{\pi}_{11}\hat{\pi}_{12}\cdots\hat{\pi}_{NN}\rangle_{\hat{\rho}_{*}}=\langle \hat{\pi}_{11}\rangle_{\hat{\rho}_{*}}\langle \hat{\pi}_{12}\rangle_{\hat{\rho}_{*}}\cdots\langle \hat{\pi}_{NN}\rangle_{\hat{\rho}_{*}}
\end{eqnarray}
where for the sake of simplicity we have omitted the explicit dependence on $H_{11},H_{12},\ldots,H_{NN}$ in all the expressions.
Now using the Eqs. \eqref{GE-1}, \eqref{GE-2} and \eqref{GE-3} one can recast \eqref{GE-6} as
\begin{eqnarray}\label{GE-7}
P(H_{11},H_{12},\ldots,H_{NN})dH_{11}dH_{12}\cdots dH_{NN}=P_{11}(H_{11})dH_{11}P_{11}(H_{12})dH_{12}\cdots P_{NN}(H_{NN})dH_{NN} \nonumber
\end{eqnarray}
Then, since $dH_{11}dH_{12}\ldots dH_{NN}$ are arbitrary small then it follows the desired result.
 \item[$(ii)$] By the Lemma \eqref{lemma1} there exist projectors $\hat{\pi}(H_{11},H_{12},\ldots,H_{NN})$, $\hat{\pi}^{\prime}(H_{11}^{\prime},H_{12}^{\prime},\ldots,H_{NN}^{\prime})$ and weak limit states $\hat{\rho}_{*}(H_{11},H_{12},\ldots,H_{NN})$, $\hat{\rho}_{*}^{\prime}(H_{11}^{\prime},H_{12}^{\prime},\ldots,H_{NN}^{\prime})$ such that
\begin{eqnarray}\label{GE-8}
&\langle \hat{\pi}\rangle_{\hat{\rho}_{*}}=P(H_{11},H_{12},\ldots,H_{NN})dH_{11}dH_{12}\cdots dH_{NN}\nonumber\\
&\nonumber\\
&\langle \hat{\pi}^{\prime}\rangle_{\hat{\rho}_{*}^{\prime}}=P(H_{11}^{\prime},H_{12}^{\prime},\ldots,H_{NN}^{\prime})dH_{11}^{\prime}dH_{12}^{\prime}\cdots dH_{NN}^{\prime}
\end{eqnarray}
where again we have omitted the explicit dependence on $H_{11},H_{12},\ldots,H_{NN}$. Since $\hat{\rho}_{*}^{\prime}$ and $\hat{\pi}^{\prime}$ refer
to the transformed density probability $P(H_{11}^{\prime},H_{12}^{\prime},\ldots,H_{NN}^{\prime})$ then it must be satisfied that
\begin{eqnarray}\label{GE-9}
&\hat{\pi}^{\prime}=\hat{U}\hat{\pi}\hat{U}^{\dag} \ \ \ \ \ \ \ \ , \ \ \ \ \ \ \ \
\hat{\rho}_{*}^{\prime}=\hat{U}\hat{\rho}_{*}\hat{U}^{\dag}
\end{eqnarray}
From Eqs. \eqref{GE-8} and \eqref{GE-9} one obtains
\begin{eqnarray}\label{GE-10}
&P(H_{11}^{\prime},H_{12}^{\prime},\ldots,H_{NN}^{\prime})dH_{11}^{\prime}dH_{12}^{\prime}\cdots dH_{NN}^{\prime}=\langle\hat{\pi}^{\prime}\rangle_{\hat{\rho}_{*}^{\prime}}=
\langle\hat{U}\hat{\pi}\hat{U}^{\dag}\rangle_{\hat{U}\hat{\rho}_{*}\hat{U}^{\dag}}=\textrm{Tr}(\hat{U}\hat{\rho}_{*}\hat{U}^{\dag}\hat{U}\hat{\pi}\hat{U}^{\dag})
=\langle\hat{\pi}\rangle_{\hat{\rho}_{*}}\nonumber\\
&=P(H_{11},H_{12},\ldots,H_{NN})dH_{11}dH_{12}\cdots dH_{NN}
\end{eqnarray}
where $\textrm{Tr}(\ldots)$ stands for the trace operation. Since the volume element $dH_{11}dH_{12}\cdots dH_{NN}$ is invariant under the transformation, i.e. $dH_{11}^{\prime}dH_{12}^{\prime}\cdots dH_{NN}^{\prime}=dH_{11}dH_{12}\cdots dH_{NN}$, then the desired result is obtained straightforwardly.
\end{itemize}
\end{proof}
\end{theorem}

\section{Physical relevance}

\subsection{Kicked rotator}
We illustrate the role played by the Gaussian ensembles in mixing quantum systems with an emblematic example of the literature: the kicked rotator \cite{stockmann, haake}.
The Hamiltonian
is given by \cite[p. 9]{stockmann}
\begin{equation}\label{kicked1}
\hat{H}=\hat{L}^{2}+\lambda \cos\hat{\theta}\sum_{m=-\infty}^{\infty} \delta(t-m\tau) \nonumber
\end{equation}
which describes the free rotation of a pendulum with angular
momentum $\hat{L}$, periodically kicked by a gravitational potential of
strength $\lambda$. The moment of inertia $I$ is
normalized to one, and $\tau$ is the kicking period.
We focus on the fully chaotic regime, that corresponds to $\lambda>5$ \cite[pp. 10, 11]{stockmann}. We show that the kicked rotator behaves like a quantum mixing system in this regime.
Let $\hat{\rho}$ be the state of the system at $t=0$
\begin{eqnarray} \label{kicked6bis}
\hat{\rho}=\sum_{k}\rho_{kk}|a_k\rangle\langle a_k|+\sum_{k\neq l}\sum_{l}\rho_{kl}|a_k\rangle\langle a_l|  \ \ \ \ \ , \ \ \ \ \ \rho_{kl}=\langle a_k|\hat{\rho}|a_l\rangle \ \ \forall k,l
\end{eqnarray}
Here $\left\{ |a_k\rangle \right\}$ is the Floquet eigenbasis \cite{stockmann}, with eigenvalues $\{e^{-i\phi _{k}}\}$ where $\{\phi_k\}$ are the so called \emph{Floquet phases}.
For an observable $\hat{O}$, after $M$ kicks one has
\begin{eqnarray} \label{kicked6}
\langle \hat{O}\rangle_{\hat{\rho}(M\tau)}=\textrm{Tr}(\hat{\rho}(M\tau)\hat{O})=\sum_{k} \rho_{kk}
O_{kk}+\sum_{k\neq l}\sum_{l\neq k}\rho_{kl}O_{kl}e^{-iM(\phi_k-\phi_l)} \ \ \ \ \ , \ \ \ \ \ O_{kl}=\langle a_k|\hat{O}|a_l\rangle \ \ \forall k,l
\end{eqnarray}
It is shown that for $\lambda>5$ the quadratic mean value of the momentum, $\langle \hat{L}^2\rangle$, exhibits exponential localization having a characteristic macroscopic width $l_s$. Moreover, if $M\gg l_s$ the phases in the factors $e^{-iM(\phi_k-\phi_l)}$ in \eqref{kicked6} oscillate rapidly
in such a way that only the diagonal terms survive; thus
\begin{equation}  \label{kicked9}
\langle \hat{O}\rangle_{\hat{\rho}(M\tau)}\simeq \sum_k \rho_{kk} O_{kk} \,\,\,\ \,\,\,\ \textrm{for} \,\,\,\ M\gg l_s
\end{equation}
Note that, if we define the diagonal part of $\hat{\rho}$ as $\hat{\rho}_{\ast}=\sum_k \rho_{kk} |a_k\rangle\langle a_k|$, then we have
\begin{eqnarray}  \label{kicked9bis}
&\langle\hat{O}\rangle_{\hat{\rho}_{\ast}}=\sum_k \rho_{kk} O_{kk}
\end{eqnarray}
Therefore we deduce that
\begin{equation}\label{kicked10}
\lim_{M\rightarrow\infty}\langle\hat{O}\rangle_{\hat{\rho}(M\tau)}=\sum_k \rho_{kk} O_{kk}=\langle\hat{O}\rangle_{\hat{\rho}_{\ast}}
\end{equation}
Recalling Eqs. \eqref{qcorrelation} and \eqref{qmixing1}, it follows that for $\lambda>5$ the kicked rotator behaves as a mixing quantum system. Even more,
$\hat{\rho}_{\ast}$ is the weak limit which is also a mixture of pure Floquet eigenstates $\{|a_k\rangle\}$ and then it can be interpreted as a decohered state, diagonal in the Floquet basis, with a decoherence time $t_D=N\tau\sim \tau l_s$.

Summing up, for $\lambda>5$ the kicked rotator is a mixing quantum system (with decoherence in the Floquet basis and induced by dephasing) and therefore,
the validity of the application of the Gaussian ensembles is justified due to Theorem \eqref{teo GE} in the classical limit $\hbar\rightarrow0$.

\subsection{Some standard approaches and the mixing quantum formalism}

Here we provide a discussion of the connection between Random matrix theory (RMT) and quantum chaotic systems, based on some attempts made in previous works (see, for instance, \cite{brody,andreev,guhr,mirlin}) and from the point of view of mixing quantum systems.
Due to the vast body of work on the subject (\cite{izrailev,bohigas,rmt2,rmt3,rmt4,rmt5,rmt6,mehta,rmt}, among others) and since our starting point are the mixing dynamical systems, we restrict the discussion to quantum systems that are chaotic in their classical limit.

Beyond the success of RMT in the prediction of statistical properties in several phenomena and its consolidation as a specific discipline,
there exist questions laying on the foundations of RMT that still remain open or partially answered. Below we quote some of these issues and discuss them from the point of view of the mixing quantum formalism.

\begin{itemize}
  \item \emph{From the point of view of the quantum mechanics, the redundant information contained in the exponentially large number of relevant periodic orbits conceals possible connections between quantum chaotic systems and RMT \cite{andreev}.}
      \vspace{0.05cm}

      In the mixing quantum formalism the only relevant information about the system is contained in the correlations between the observables and the weak limit, as shown in Eqs.~\eqref{qcorrelation} and \eqref{qmixing1}. As we mentioned at the beginning of the Section 4, the weak limit process is a type of decoherence of the observables \cite{0,vhove,Omnes}. Thus, the redundant information is suppressed by the cancellation of the quantum mixing correlations in the asymptotic limit.

  \item \emph{Although there have been several ways to deduce the BGS conjecture, for instance the non--linear $\sigma$--model \cite{mirlin} and the semiclassical trace formula \cite{stockmann}, all
  the attempts are based on some kind of semiclassical approximation \cite{guhr}.}
     \vspace{0.05cm}

     One of the advantages of the mixing quantum approach is that it allows to deduce the Gaussian ensembles distributions as a consequence of the quantum factorization property in terms of operators and states (Theorem \ref{teo GE}), i.e. in the language of the quantum mechanical operators. However, the classical limit has to be considered in order to apply the quantum factorization property.

  \item \emph{In Gaussian ensembles the behavior is studied along the energy axis rather than the time axis, while the thermodynamics systems evolve along the time axis. Furthermore, since there is no way of describing mathematically the transition from one level to the next then there is no analog of the time arrow of thermodynamics \cite{brody}.}
      \vspace{0.05cm}
      From Eqs. \eqref{GE-1} and \eqref{GE-2} it can be seen that the evolution is involved (in the asymptotic limit) since the joint density probability $P(H_{11},H_{12},\ldots,H_{NN})$ and its marginals $P_{11}(H_{11}),P_{12}(H_{12}),\ldots,P_{NN}(H_{NN})$ are expressed in terms of traces of projectors in the weak limit, which also allows to give a probabilistic interpretation according to the Born rule. In addition, one can say that in a mixing quantum system the time arrow is due to quantum mixing correlations which gives place to an irreversible dynamics expressed
      by the mixture character of the weak limit, as shown in Eqs. \eqref{kicked9bis} and \eqref{kicked10}.

  \item \emph{Since the trajectories of a dynamical system are system--specific, the role of the ensemble theory of statistical mechanics rules out. Instead, ensembles of different Hamiltonians are able to mimic the statistical behavior of a dynamical system.}
      \vspace{0.05cm}

      In the mixing quantum formalism the key point is to consider the statistical description given by the Ergodic Hierarchy in terms of correlations between subsets in phase space instead of using trajectories. In particular, for the mixing level one has that any two subsets separated enough in time have a null correlation and can be interpreted as statistically independent events. Moreover, this statistical independence property can be generalized for subsets in a sequence, the factorization property (Lemma \ref{lemma1}), that characterizes randomness between subsets.
In turn, in the classical limit the quantum factorization property (Theorem \ref{theorem}) allows to express the randomness in terms of the factorization of mean values in the weak limit.

\end{itemize}

In order to illustrate how the Gaussian ensembles are deduced from the standard approaches in RMT, below we provide a schematic picture showing some of them along with the mixing quantum formalism.

\begin{eqnarray}
& \textrm{\textbf{Wigner--Dyson original approach}} \nonumber\\
&\textrm{universality in local fluctuations of quantum spectra} \ \Longrightarrow \ \textrm{randomness and invariance} \  \Longrightarrow  \ \textrm{Gaussian ensembles} \nonumber
& \nonumber\\
& \nonumber\\
& \textrm{\textbf{semiclassical trace formula}} \nonumber\\
&\textrm{Gutzwiller trace formula} \ \Longrightarrow \ n-\textrm{point correlation functions in the classical limit} \  \Longrightarrow  \ \textrm{Gaussian ensembles} \nonumber
& \nonumber\\
& \nonumber\\
&\textrm{\textbf{mixing quantum formalism}} \nonumber\\
&\textrm{factorization property} \ \Longrightarrow \  \textrm{quantum factorization property in the classical limit} \ \Longrightarrow  \ \textrm{Gaussian ensembles} \nonumber
\end{eqnarray}

\section{Conclusions}
We have proposed a novel way to deduce the Gaussian ensembles within the quantum mixing level of the quantum ergodic hierarchy. The relevance of our main contribution, Theorems \ref{theorem} and \ref{teo GE}, lies in the following remarks:
\begin{itemize}
  \item In the classical limit the randomness condition of Gaussian ensembles results as a consequence of the quantum mixing correlations.
  \item The probability density function for the Hamiltonian matrix elements can be computed in terms of the mean value of a projector in a weak limit (Eq. \eqref{GE-2}). In addition, this can be considered as a kind of analog of the Born rule.
\item For the kicked rotator case we show that mixing quantum formalism links decoherence, in the Floquet basis and induced by dephasing, with Gaussian ensembles in terms of the weak limit which also can be interpreted as a decohered state. Moreover, starting with a pure state the mixture character of its weak limit expresses the irreversible dynamics of the mixing quantum systems, as shown in Eqs. \eqref{kicked9bis} and \eqref{kicked10}.
\item Going further, from Theorems \ref{theorem} and \ref{teo GE} we could rephrase the statement of the Bohigas--Giannoni--Schmit conjecture \cite{Boh84} for the family of mixing quantum systems as: \emph{Hamiltonian matrix elements of mixing quantum systems show, in the classical limit, the same probability density function as predicted by Gaussian ensembles}.

\end{itemize}

Summarizing, we conclude that the ``imitation" of statistical properties of quantum systems having a mixing (and therefore chaotic) classical limit arises as a consequence of the quantum factorization property within the mixing quantum formalism.

\section*{Acknowledgments}
This work was partially supported by CONICET and Universidad Nacional de La Plata, Argentina.

\appendix
\section{Proof of Lemma 3.1}
\begin{proof}
Let us write $f_{*}(q,p)$ as a linear combination of characteristic functions in the form $f_{*}(q,p)=\sum_i \alpha_i1_{C_i}(q,p)$ with
$C_i\cap C_{j}=\emptyset$ if $i\neq j$ and $\bigcup_{i}C_i=\Gamma$. Then $\int_{\Gamma}f_{\ast}(q,p)dqdp=\sum_i \alpha_i\mu(C_i)=1$.
For $A_1,A_2\subseteq\Gamma$, from definition \eqref{correlation} we can write
\begin{equation}\label{lemma1-5}
\mu(T_tA_1\cap A_2)=C(T_tA_1,A_2)+\mu(A_1)\mu(A_2)
\end{equation}
Let us compute the following expression:
\begin{eqnarray}\label{lemma1-6}
&\sum_i \alpha_i \sum_j \alpha_j\mu(T_tC_i \cap (A_1 \cap C_j \cap A_2))=
\sum_{i} \alpha_i \sum_{j}\alpha_j\left[C(T_tC_i,A_1 \cap C_j \cap A_2)+\mu(C_i)\mu(C_j\cap A_1 \cap A_2)\right]\nonumber\\
&=\sum_i \alpha_i \sum_j \alpha_j C(T_tC_i,A_1 \cap C_j \cap A_2)+\sum_i \alpha_i \mu(C_i)\sum_j \alpha_j \mu(C_j\cap A_1 \cap A_2)\nonumber\\
&=\sum_{i} \alpha_i \sum_{j}\alpha_j C(T_tC_i,A_1 \cap C_j \cap A_2)+\sum_j \alpha_j \int_{\Gamma}1_{C_j \cap A_1 \cap A_2}(q,p)dqdp\nonumber\\
&=\sum_{i}\alpha_i \sum_{j}\alpha_j C(T_tC_i,A_1 \cap C_j \cap A_2)+\int_{\Gamma}\sum_j \alpha_j 1_{C_j}(q,p) 1_{A_1}(q,p) 1_{A_2}(q,p)dqdp\nonumber\\
&=\sum_{i}\alpha_i \sum_{j}\alpha_j C(T_tC_i,A_1 \cap C_j \cap A_2)+\int_{\Gamma}f_{\ast}(q,p) 1_{A_1}(q,p) 1_{A_2}(q,p)dqdp
\end{eqnarray}
Also we have that
\begin{eqnarray}\label{lemma1-7}
&\sum_i \alpha_i \sum_j \alpha_j\mu(T_tC_i \cap A_1 \cap C_j \cap A_2)=\sum_i \alpha_i \sum_j \alpha_j\mu(T_tC_i \cap T_t(T_{-t}A_1) \cap C_j \cap A_2)\nonumber\\
&=\sum_i \alpha_i \sum_j \alpha_jC(T_t(C_i \cap T_{-t}A_1), C_j \cap A_2)+\sum_i \alpha_i \mu(C_i \cap T_{-t}A_1)\sum_j \alpha_j\mu(C_j\cap A_2)\nonumber\\
&=\sum_i \alpha_i \sum_j \alpha_jC(T_t(C_i \cap T_{-t}A_1), C_j \cap A_2)+\sum_i \alpha_i \int_{\Gamma}1_{C_i}(q,p) 1_{T_{-t}A_1}(q,p)dqdp\sum_j \alpha_j \int_{\Gamma}1_{C_j}(q,p)1_{A_2}(q,p)dqdp\nonumber\\
&=\sum_i \alpha_i \sum_j \alpha_jC(T_t(C_i \cap T_{-t}A_1), C_j \cap A_2)+\int_{\Gamma}f_{\ast}(q,p) 1_{T_{-t}A_1}(q,p)dqdp\int_{\Gamma}f_{\ast}(q,p)1_{A_2}(q,p)dqdp\nonumber\\
&=\sum_i \alpha_i \sum_j \alpha_jC(T_t(C_i \cap T_{-t}A_1), C_j \cap A_2)+\int_{T_{-t}A_1}f_{\ast}(q,p)dqdp\int_{\Gamma}f_{\ast}(q,p)1_{A_2}(q,p)dqdp
\end{eqnarray}
Now by the definition of the Frobenius--Perron operator $P_t$ and since $f_{\ast}$ is a fixed point of $P_t$, we have
\begin{equation}\label{lemma1-8}
\begin{split}
\int_{T_{-t}A_1}f_{\ast}(q,p)dqdp=\int_{A_1}P_tf_{\ast}(q,p)dqdp=\int_{A_1}f_{\ast}(q,p)dqdp=\int_{\Gamma}f_{\ast}(q,p)1_{A_1}(q,p)dqdp
\end{split}
\end{equation}
Then using \eqref{lemma1-8} we can recast \eqref{lemma1-7} as
\begin{eqnarray}\label{lemma1-9}
&\sum_i \alpha_i \sum_j \alpha_j\mu(T_tC_i \cap A_1 \cap C_j \cap A_2)\nonumber\\
&=\sum_i \alpha_i \sum_j \alpha_jC(T_t(C_i \cap T_{-t}A_1), C_j \cap A_2)+\int_{\Gamma}f_{\ast}(q,p)1_{A_1}(q,p)dqdp\int_{\Gamma}f_{\ast}(q,p)1_{A_2}(q,p)dqdp
\end{eqnarray}
In the limit $t\rightarrow\infty$, the correlations $C(T_tC_i,A_1 \cap C_j \cap A_2)$ and $C(T_t(C_i \cap T_{-t}A_1), C_j \cap A_2)$ become vanishingly small
due to Eq. (\ref{mixing1}) since $T_t(C_i \cap T_{-t}A_1)\subseteq T_tC_i$ and $C_j \cap A_2$ are sufficiently separated in time for large times. Therefore, from Eqs. \eqref{lemma1-6} and \eqref{lemma1-9} we have
\begin{equation}\label{lemma1-10}
\begin{split}
&\lim_{t\rightarrow\infty}\sum_i \alpha_i \sum_j \alpha_j\mu(T_tC_i \cap A_1 \cap C_j \cap A_2)\nonumber\\
&=\int_{\Gamma}f_{*}(q,p)1_{A_1}(q,p)1_{A_2}(q,p)dqdp=\int_{\Gamma}f_{*}(q,p)1_{A_1}(q,p)dqdp\int_{\Gamma}f_{*}(q,p)1_{A_2}(q,p)dqdp
\end{split}
\end{equation}
If we have $n$ characteristic functions $1_{A_1},1_{A_2},\ldots,1_{A_n}$ we can apply the last equality $n-1$ times so that we prove the desired result.
\end{proof}
\section{Proof of Lemma 4.1}
\begin{proof}
Let $s$ be a real number and let us consider the evolved operator $\hat{U}_s^{\dag}\hat{O}\hat{U}_s$ for a given operator $\hat{O}$. From Eq. \eqref{qmixing1} we have
\begin{equation}\label{qmixing2}
\begin{split}
\lim_{t\rightarrow\infty}\langle\hat{U}_{s}^{\dagger}\hat{O}\hat{U}_{s}\rangle_{\hat{\rho}(t)}-\langle\hat{U}_{s}^{\dagger}\hat{O}\hat{U}_{s}\rangle_{\hat{\rho}_{*}}=0
\end{split}
\end{equation}
and applying trace properties we can rewrite it as
\begin{equation}\label{qmixing3}
\begin{split}
\lim_{t\rightarrow\infty}\langle\hat{O}\rangle_{\hat{\rho}(t+s)}-\langle\hat{O}\rangle_{\hat{U}_{s}\hat{\rho}_{*}\hat{U}_{s}^{\dagger}}=0
\end{split}
\end{equation}
Since
\begin{equation}\label{qmixing4}
\begin{split}
\lim_{t\rightarrow\infty}\langle\hat{O}\rangle_{\hat{\rho}(t+s)}=\lim_{t\rightarrow\infty}\langle\hat{O}\rangle_{\hat{\rho}(t)}=
\langle\hat{O}\rangle_{\hat{\rho}_{*}}
\end{split}
\end{equation}
then it follows that
$\langle\hat{O}\rangle_{\hat{U}_{s}\hat{\rho}_{*}\hat{U}_{s}^{\dagger}}=\langle\hat{O}\rangle_{\hat{\rho}_{*}}$ for all observable $\hat{O}$, which means that
\begin{equation}\label{lemma1-17}
\begin{split}
\hat{U}_s\hat{\rho}_{\ast}\hat{U}_s^{\dag}=\hat{\rho}_{\ast} \ \ \ \ \ \ \forall s\in\mathbb{R}
\end{split}
\end{equation}
\end{proof}
\section{Proof of Lemma 4.2}
\begin{proof}
From the definition of the Weyl symbol, Eq. \eqref{wigner1}, one has
\begin{eqnarray}\label{prop1}
\widetilde{W}_{\hat{A}}(q,p)=\int_{\mathbb{R}}\langle q+\Delta| \hat{A}|q-\Delta\rangle e^{2i\frac{p\Delta}{\hbar}}d\Delta
\end{eqnarray}
Then it follows that
\begin{eqnarray}\label{prop2}
\widetilde{W}_{\hat{A}}(T_tq,T_tp)=\int_{\mathbb{R}}\langle T_tq+\Delta| \hat{A}|T_tq-\Delta\rangle e^{2i\frac{T_tp\Delta}{\hbar}}d\Delta
\end{eqnarray}
Now we make the change of variables $\Delta\longrightarrow \widetilde{\Delta}=T_{-t}\Delta$, then
\begin{eqnarray}\label{prop3}
\Delta=T_t\widetilde{\Delta}  \ \ \ \ \textrm{and} \ \ \ \ \
d\Delta=|T_t|d\widetilde{\Delta}
\end{eqnarray}
being $|T_t|$ the Jacobian determinant of $T_t$ restricted to the coordinates $q$.
Moreover, since the Liouville classical evolution preserves the volume of phase space we can assume that $|T_t|=1$.
Then, using \eqref{prop3} and that $|T_t|=1$ we can recast \eqref{prop2} as
\begin{eqnarray}\label{prop4}
\widetilde{W}_{\hat{A}}(T_tq,T_tp)=\int_{\mathbb{R}}\langle T_tq+T_t\widetilde{\Delta}| \hat{A}|T_tq-T_t\widetilde{\Delta}\rangle e^{2i\frac{T_tpT_t\widetilde{\Delta}}{\hbar}}d\widetilde{\Delta}
\end{eqnarray}
It is clear that
\begin{eqnarray}\label{prop5}
&\langle T_tq+T_t\widetilde{\Delta}|=\langle T_t(q+\widetilde{\Delta})|=\langle q+\widetilde{\Delta}|\hat{U}^{\dag}(t)\nonumber\\
&|T_tq-T_t\widetilde{\Delta}\rangle=|T_t(q-\widetilde{\Delta})\rangle=\hat{U}(t)|q-\widetilde{\Delta}\rangle
\end{eqnarray}
and also
\begin{eqnarray}\label{prop6}
&e^{2i\frac{T_tpT_t\widetilde{\Delta}}{\hbar}}=e^{2i\frac{p\widetilde{\Delta}}{\hbar}}\Longleftrightarrow
\frac{T_tpT_t\widetilde{\Delta}}{\hbar}-\frac{p\widetilde{\Delta}}{\hbar}=m\pi \ \ \ \textrm{with} \ \ \ m\in\mathbb{Z} \ \ \ \Longleftrightarrow \ \ \ p(t)\widetilde{\Delta}(t)-p\widetilde{\Delta}=mh/2
\end{eqnarray}
Since the quantum phase space is grained due to the Indetermination Principle by cells of volume $h/2$, then in the classical limit for $h$ vanishingly small
the condition of \eqref{prop6} is satisfied.
Therefore, replacing \eqref{prop5} and \eqref{prop6} in \eqref{prop4} we obtain the desired result.
\end{proof}
\section{Proof of Lemma 4.3}
\begin{proof}
By applying the definition of Frobenius-Perron operator, Eq. \eqref{perron}, to the Wigner function $W_{\hat{\rho}_{\ast}}(q,p)$, using Lemma \ref{lemma2} and Eq. \eqref{wigner8}, we have
\begin{equation}\label{lemma1-18}
\int_{A}P_t W_{\hat{\rho}_{\ast}}(q,p)dqdp=\int_{T_{-t}A}W_{\hat{\rho}_{\ast}}(q,p)dqdp=
\int_{\mathbb{R}^2}W_{\hat{\rho}_{\ast}}(q,p)1_{T_{-t}A}(q,p)dqdp
\end{equation}
Now let $\hat{\pi}_A$ be the operator such that $\widetilde{W}_{\hat{\pi}_A}(q,p)=1_{A}(q,p)$. By applying Eq. \eqref{lemaweyl1} to $\hat{\pi}_{A}$ it follows that $\widetilde{W}_{\hat{U}_t^{\dag}\hat{\pi}_{A}\hat{U}_t}(q,p)=1_{A}(T_tq,T_tp)=1_{T_{-t}A}(q,p)$. Then using this in Eq. \eqref{lemma1-18} we have
\begin{eqnarray}\label{lemma1-18bis}
&\int_{A}P_tW_{\hat{\rho}_{\ast}}(q,p)dqdp=\int_{\mathbb{R}^2}W_{\hat{\rho}_{\ast}}(q,p)\widetilde{W}_{\hat{U}_t^{\dag}\hat{I}_{A}\hat{U}_t}(q,p)dqdp\nonumber\\
&=\textrm{Tr}(\hat{\rho}_{\ast}\hat{U}_t^{\dag}\hat{I}_{A}\hat{U}_t)=\textrm{Tr}(\hat{U}_t\hat{\rho}_{\ast}\hat{U}_t^{\dag}\hat{I}_{A})=
\textrm{Tr}(\hat{\rho}_{\ast}\hat{I}_{A})
=\int_{\mathbb{R}^2}W_{\hat{\rho}_{\ast}}(q,p)\widetilde{W}_{{I}_{A}}(q,p)dqdp=
\int_{\mathbb{R}^2}W_{\hat{\rho}_{\ast}}(q,p)1_A(q,p)dqdp\nonumber\\
&=\int_{A}W_{\hat{\rho}_{\ast}}(q,p)dqdp
\end{eqnarray}
Then, since $A\subseteq\Gamma$ is arbitrary and given that $P_tW_{\hat{\rho}_{\ast}}(q,p)$ and $W_{\hat{\rho}_{\ast}}(q,p)$ are non negative, it follows that
$P_t W_{\hat{\rho}_{\ast}}(q,p)=W_{\hat{\rho}_{\ast}}(q,p)$ almost everywhere on $\mathbb{R}^2$. Nevertheless, since we only use
$P_t W_{\hat{\rho}_{\ast}}(q,p)$ and $W_{\hat{\rho}_{\ast}}(q,p)$ by means of integrals we can consider without loss of generality that
$P_t W_{\hat{\rho}_{\ast}}(q,p)=W_{\hat{\rho}_{\ast}}(q,p)$ for all $(q,p)\in\mathbb{R}^2$. This completes the proof.
\end{proof}
\section{Proof of Lemma 5.1}
\begin{proof}
Since $P(H_{11},H_{12},\ldots,H_{NN})$, $P_{11}(H_{11})$, $P_{12}(H_{12})$, $\ldots$, $P_{NN}(H_{NN})$ are positive numbers then one can consider $dH_{11}$, $dH_{12}$, $\ldots$, $dH_{NN}>0$ sufficiently small such that
\begin{eqnarray}\label{demo 5.1-1}
&0<P_{ij}(H_{ij})dH_{ij}<1/2 \ \ \ \ \ \ \ \ \forall \ i,j=1,\ldots,N\nonumber\\
& \nonumber\\
&0<P(H_{11},H_{12},\ldots,H_{NN})dH_{11}dH_{12}\cdots dH_{NN}<1/2
\end{eqnarray}
Let $\alpha,\beta$ be positive numbers such that
\begin{eqnarray}\label{demo 5.1-2}
&\alpha<\textrm{min}\left\{P(H_{11},H_{12},\ldots,H_{NN}),P_{11}(H_{11}), P_{12}(H_{12}),\ldots, P_{NN}(H_{NN})\right\} \nonumber \\
& \\ \nonumber
&\textrm{max}\left\{P(H_{11},H_{12},\ldots,H_{NN}),P_{11}(H_{11}), P_{12}(H_{12}),\ldots, P_{NN}(H_{NN})\right\}<\beta\leq 1/2
\end{eqnarray}
Since $0<\alpha<\beta\leq 1/2$ then there exists $\gamma\geq0$ such that $\alpha+\beta+\gamma=1$.
From eqs. \eqref{demo 5.1-2} one has
\begin{eqnarray}\label{demo 5.1-3}
\sqrt{\frac{P(H_{ij})dH_{ij}}{\alpha}}>1   \ \ \ \ \ \ \ \ , \ \ \ \ \ \ \ \  \sqrt{\frac{P(H_{ij})dH_{ij}}{\beta}}<1 \ \ \ \ \ \ \ \ \ \ \ \ \ \ \ \ \forall \ i,j=1,\ldots,N
\end{eqnarray}

and

\begin{eqnarray}\label{demo 5.1-4}
\sqrt{\frac{P(H_{11},H_{12},\ldots,H_{NN})dH_{11}dH_{12}\cdots dH_{NN}}{\alpha}}>1   \ \ \ \  ,  \ \ \ \  \sqrt{\frac{P(H_{11},H_{12},\ldots,H_{NN})dH_{11}dH_{12}\cdots dH_{NN}}{\beta}}<1
\end{eqnarray}
Now consider the systems of equations

\begin{eqnarray}\label{demo 5.1-5}
\left\{\begin{array}{cc}  u_{ij}^2+v_{ij}^2=1 \\
\\
                 \left(\frac{u_{ij}}{\sqrt{\frac{P(H_{ij})dH_{ij}}{\alpha}}}\right)^2+\left(\frac{v_{ij}}{\sqrt{\frac{P(H_{ij})dH_{ij}}{\beta}}}\right)^2=1  \end{array}\right.
\end{eqnarray}
and
\begin{eqnarray}\label{demo 5.1-6}
\left\{\begin{array}{cc}  u^2+v^2=1 \\
\\
                 \left(\frac{u}{\sqrt{\frac{P(H_{11},H_{12},\ldots,H_{NN})dH_{11}dH_{12}\cdots dH_{NN}}{\alpha}}}\right)^2+\left(\frac{v}{\sqrt{\frac{P(H_{11},H_{12},\ldots,H_{NN})dH_{11}dH_{12}\cdots dH_{NN}}{\beta}}}\right)^2=1  \end{array}\right.
\end{eqnarray}
Eqs. \eqref{demo 5.1-5} and \eqref{demo 5.1-6} represent the intersection of the unitary circle with ellipses whose major axis are equal to $\sqrt{\frac{P(H_{ij})dH_{ij}}{\alpha}}$, $\sqrt{\frac{P(H_{11},H_{12},\ldots,H_{NN})dH_{11}dH_{12}\cdots dH_{NN}}{\alpha}}$ and whose minor axis are equal to $\sqrt{\frac{P(H_{ij})dH_{ij}}{\beta}}$, $\sqrt{\frac{P(H_{11},H_{12},\ldots,H_{NN})dH_{11}dH_{12}\cdots dH_{NN}}{\beta}}$ for all $i,j=1,\ldots,N$.
Then it follows that \eqref{demo 5.1-5} and \eqref{demo 5.1-6} have solutions $(u_{ij},v_{ij})$, $(u,v)$ with $u_{ij},v_{ij},u,v\neq0$ for all $i,j=1,\ldots,N$.
Let $\{|\psi_i\rangle\}_{i=1}^N$ be the eigenbasis of the Hamiltonian $\hat{H}$.
Now since $\alpha+\beta+\gamma=1$ then one can define the state
\begin{eqnarray}\label{demo 5.1-7}
\hat{\rho}_{*}=\alpha |\psi_1\rangle\ \langle\psi_1|+\beta |\psi_2\rangle\ \langle\psi_2|+\gamma |\psi_3\rangle\ \langle\psi_3|
\end{eqnarray}
and the operators
\begin{eqnarray}\label{demo 5.1-8}
&\hat{\pi}_{ij}= \left( u_{ij} |\psi_1\rangle\ + v_{ij} |\psi_2\rangle\ \right)\left( u_{ij} \langle\psi_1| + v_{ij} \langle\psi_2| \right) \ \ \ \ \ \ \ \ \ \ \ \ \ \ \ \ \forall \ i,j=1,\ldots,N \nonumber \\
& \nonumber\\
&\hat{\pi}= \left( u |\psi_1\rangle\ + v |\psi_2\rangle\ \right)\left( u \langle\psi_1| + v \langle\psi_2| \right)
\end{eqnarray}
Since $u_{ij}^2+v_{ij}^2=1$ , $u^2+v^2=1$ and due to the eigenbasis $\{|\psi_i\rangle\}_{i=1}^N$ is orthonormal then one has that $\hat{\pi}_{ij}$ and $\hat{\pi}$ are projectors for all $i,j=1,\ldots,N$.
From Eqs. \eqref{demo 5.1-5} and \eqref{demo 5.1-6} one obtains
\begin{eqnarray}\label{demo 5.1-9}
&\langle \pi_{ij} \rangle_{\hat{\rho}_{*}}=\textrm{Tr}(\hat{\rho}_{*}\pi_{ij})=\alpha u_{ij}^2+\beta v_{ij}^2=P(H_{ij})dH_{ij} \ \ \ \ \ \ \ \ \ \ \ \ \ \ \ \ \forall \ i,j=1,\ldots,N \nonumber \\
&\nonumber\\
&\langle \pi \rangle_{\hat{\rho}_{*}}=\textrm{Tr}(\hat{\rho}_{*}\pi)=\alpha u^2+\beta v^2=P(H_{11},H_{12},\ldots,H_{NN})dH_{11}dH_{12}\cdots dH_{NN}
\end{eqnarray}
where $\textrm{Tr}(\ldots)$ denotes the trace operation.

Due to eq. \eqref{demo 5.1-9} and since $dH_{11}dH_{12}\ldots dH_{NN}$ can be taken arbitrary small
one can see that the projector $\hat{\pi}_{ij}$ is associated with the probability of the $ij$--th Hamiltonian matrix element is $H_{ij}$ and $\hat{\pi}$ is associated with the joint probability of the Hamiltonian matrix elements are $H_{11},H_{12},\ldots,H_{NN}$. Physically, this can be considered a kind of analog of the Born rule.
Thus, it is reasonable to consider that there exist a relationship between $\hat{\pi}$ and $\hat{\pi}_{ij}$ for all $i,j=1,\ldots,N$.
Assuming that $\hat{\pi}$ is an analytical function of the projectors $\pi_{ij}$ one has
\begin{eqnarray}\label{demo 5.1-10}
\hat{\pi}=\sum_{k_{11},k_{12},\ldots,k_{NN}=0}^{\infty}a_{k_{11},k_{12},\ldots,k_{NN}}(\hat{\pi}_{11})^{k_{11}}(\hat{\pi}_{12})^{k_{12}}\cdots(\hat{\pi}_{NN})^{k_{NN}}
\end{eqnarray}
where $a_{k_{11},k_{12},\ldots,k_{NN}}$ are constant coefficients. Since the trace $\langle \pi \rangle_{\hat{\rho}_{*}}$ is proportional to the joint density probability $P(H_{11},H_{12},\ldots,H_{NN})$ then all the projectors $\hat{\pi}_{ij}$ must be appear on the product $(\hat{\pi}_{11})^{k_{11}}(\hat{\pi}_{12})^{k_{12}}\cdots(\hat{\pi}_{NN})^{k_{NN}}$ in eq. \eqref{demo 5.1-10}, i.e. $k_{ij}\neq0$ for all $i,j=1,\ldots,N$. Moreover, using that $\pi_{ij}$ is a projector for all $i=1,\ldots,N$ then the only power that survive in \eqref{demo 5.1-10} is $k_{ij}=1$ for all $i,j=1,\ldots,N$, thus one can recast \eqref{demo 5.1-10} as
\begin{eqnarray}\label{demo 5.1-11}
\hat{\pi}=K \hat{\pi}_{11}\hat{\pi}_{12}\cdots\hat{\pi}_{NN}
\end{eqnarray}
where $K$ is a constant coefficient to be determined by the condition of $\hat{\pi}$ is a projector. Indeed, since $\hat{\pi}^2=\hat{\pi}$ one has
\begin{eqnarray}\label{demo 5.1-11}
K^2 (\hat{\pi}_{11}\hat{\pi}_{12}\cdots\hat{\pi}_{NN})^2=K \hat{\pi}_{11}\hat{\pi}_{12}\cdots\hat{\pi}_{NN}
\end{eqnarray}
Now by taking the classical limit $\hbar\rightarrow0$ the product $\hat{\pi}_{11}\hat{\pi}_{12}\cdots\hat{\pi}_{NN}$ becomes commutative so
\begin{eqnarray}\label{demo 5.1-12}
(\hat{\pi}_{11}\hat{\pi}_{12}\cdots\hat{\pi}_{NN})^2=\hat{\pi}_{11}^2\hat{\pi}_{12}^2\cdots\hat{\pi}_{NN}^2=\hat{\pi}_{11}\hat{\pi}_{12}\cdots\hat{\pi}_{NN}
\end{eqnarray}
By replacing this in eq. \eqref{demo 5.1-11} one obtains $K^2=K$ and since $\hat{\pi}$ cannot be the null projector it follows that $K=1$. Therefore, $\hat{\pi}=\hat{\pi}_{11}\hat{\pi}_{12}\cdots\hat{\pi}_{NN}$.
  \end{proof}

\section*{References}

\begin{thebibliography}{99}
\bibitem{Wigner gaussian} E. Wigner,
\textit{Ann. of Math.} {\bf 62}, 548--564 (1955).

\bibitem{Dyson} Freeman J. Dyson,
\textit{J. Math. Phys.} {\bf 3}, 1199 (1962).

\bibitem{Boh84} O. Bohigas, M. J. Giannoni and C. Schmit,
\textit{Phys. Rev. Lett.} {\bf 52}, 1 (1984).

\bibitem{apps1} L. C. Garc\'{i}a del Molino, K. Pakdaman, J. Touboul, G. Wainrib,
\textit{Phys. Rev. E} {\bf 88}, 042824 (2013).

\bibitem{apps2} K. Rajan, L. F. Abbott,
\textit{Phys. Rev. Lett.} {\bf 97}, 188104 (2006).

\bibitem{apps3} M. Schreibera, U. Grimma, R. A. R\"{o}mer, J. Zhonga, \textit{Physica A } {\bf 266}, 477-480 (1999).

\bibitem{apps4} P. Shukla, \textit{Physica A} {\bf 288}, 119-129 (2000).

\bibitem{0} M. Castagnino and O. Lombardi,
\textit{Physica A} {\bf 388}, 247--267 (2009).

\bibitem{NACHOSKY MARIO} I. Gomez and M. Castagnino,
\textit{Physica A} {\bf 393}, 112--131 (2014).

\bibitem{casati model}
M. Castagnino,
\textit{Phys. Lett. A} {\bf 357}, 97-100 (2006).

\bibitem{casati verdadero}
G. Casati and T. Prosen,
\textit{Phys. Lett. A} {\bf 72}, 032111 (2005).


\bibitem{stockmann}
H. Stockmann,
\textit{Quantum Chaos - An Introduction}, Cambridge Univ. Press, Cambridge (1999).

\bibitem{haake}
F. Haake,
\textit{Quantum Signature of Chaos}, 2nd edition, Springer-Verlag, Heidelberg (2001).


\bibitem{vhove} L. van Hove,
\textit{Physica A} \textbf{20}, 603 (1954).

\bibitem{Omnes} R. Omn\`{e}s,
\textit{The Interpretation of Quantum Mechanics}, Princeton University, Princeton (1994).


\bibitem{Wigner} M. Hillery, R. O'Connell, M. Scully and E. Wigner,
\textit{Phys. Rep}. \textbf{106}, 121-167 (1984).

\bibitem{Symb} G. Dito and D. Sternheimer,
\textit{IRMA Lectures in Mathematics and Theoretical Physics} {\bf 1}, 9-54 (2002).


\bibitem{casati libro} G. Casati, B. Chirikov, \emph{Quantum Chaos: between order and disorder}, Cambridge
University Press, Cambridge (1995).

\bibitem{chirikov izrailev} B. Chirikov, F. Izrailev, D. Shepelyansky, \emph{Physica A}, {\bf 33}, 77--88 (1988).

\bibitem{var} V. Varadarajan, \textit{Geometry of Quantum Theory}, Springer Verlag, New York (1970).

\bibitem{brody} T. Brody, J. Flores, J. French, P. Mello, A. Pandey, S. Wong, \textit{Rev. Mod. Phys.}, {\bf 53}, 3 (1981).

\bibitem{andreev} A. Andreev, O. Agam, B. Simons, B. Altshuler, \textit{Phys. Rev. Lett.} {\bf 76}, 21 (1996).


\bibitem{guhr} T. Guhr, A. Mull\"{e}r--Groeling, H Weidenmu\"{u}ller, \textit{Phys. Rep.}, {\bf 299}, 189--425 (1998).

\bibitem{mirlin} A. Mirlin, \textit{Phys. Rep.}, {\bf 326}, 259--382 (2000).

\bibitem{izrailev} F. Izrailev, \textit{Phys. Rep.}, {\bf 196}, 299--392 (1990).

\bibitem{bohigas} O. Bohigas, \textit{Random Matrix Theories And Chaotic Dynamics}, Les Houches, Session LII, Chaos et Physique Quantique/Chaos and Quantum Physics, Elsevier Science Publishers B. V. (1991).

\bibitem{rmt2} O. Bohigas, S. Tomsovic, D. Ullmo, \textit{Phys. Rep.}, {\bf 223}, 43--133 (1993).

\bibitem{rmt3} A. Kudrolli, S. Sridhar, A. Pandey, R. Ramaswamy, \textit{Phys. Rev. E}, {\bf 49}, 1 (1994).

\bibitem{rmt4} Y. Shimizu, A. Shudo, \textit{Chaos, Solitons and Fractals}, {\bf 5}, 7, 1337--1362 (1995).

\bibitem{rmt5} V. Zelevinsky, B. Brown, N. Frazier, M. Horoi, \textit{Phys. Rep.}, {\bf 276}, 85--176 (1996).

\bibitem{rmt6} M. Caselle, U. Magnea, \textit{Phys. Rep.}, {\bf 394}, 41--156 (2004).

\bibitem{mehta} M. Mehta, \textit{Random Matrices}, Volume 142, 3rd Edition (Pure and Applied Mathematics Series), Academic Press, North America (2004).

\bibitem{rmt} G. Anderson, A. Guionnet, O. Zeitouni, \textit{An Introduction to Random Matrices}, Cambridge studies in advanced mathematics, Cambridge
University Press, Cambridge (2009).

\end{thebibliography}
\end{document}